\documentclass{soups}

\usepackage{times}
\usepackage{amssymb}
\usepackage{graphicx}
\usepackage{bmpsize}
\usepackage{epsfig}
\usepackage{latexsym}
\usepackage{amsfonts}
\usepackage{eucal}
\usepackage{amsmath}
\usepackage{algorithmic}
\usepackage{algorithm}
\usepackage{stfloats}
\usepackage{fixltx2e}
\usepackage{setspace}
\usepackage[nodots]{numcompress}
\usepackage{array}
\usepackage{caption}
\usepackage{subcaption}

\begin{document}

\title{An HMM-based Behavior Modeling Approach for Continuous Mobile Authentication\titlenote{This paper was published as \cite{roy2014hmm}}}

\author{
\alignauthor
Aditi Roy, Tzipora Halevi and Nasir Memon\\
\affaddr{New York University, Polytechnic School of Engineering, Brooklyn, NY, USA}\\
\email {ar3824@nyu.edu, thalevi@nyu.edu, memon@nyu.edu}}

\maketitle

\begin{abstract}
This paper studies continuous authentication for touch interface
based mobile devices. A Hidden Markov Model (HMM) based behavioral
template training approach is presented, which does not require
training data from other subjects other than the owner of the
mobile. The stroke patterns of a user are modeled using a
continuous left-right HMM. The approach models the horizontal and
vertical scrolling patterns of a user since these are the basic
and mostly used interactions on a mobile device. The effectiveness
of the proposed method is evaluated through extensive experiments
using the Touchalytics database which comprises of touch data over
time. The results show that the performance of the proposed
approach is better than the state-of-the-art method.

\end{abstract}


\section{Introduction}
\label{sec:intro}

Existing technology typically requires users to
authenticate themselves based on passwords, which have been shown
to be vulnerable to various attacks, including password guessing
and eavesdropping (\cite{Jo12,data12}).

Multiple methods were proposed to replace text passwords with
graphical passwords \cite{DP00, DMR04, AD04, WK04, Birget06,
DMB07}. With the growing popularity of touch interface based
mobile devices, the touch-surface has become the dominant
human-computer interface. This has led to the need for
authentication techniques better suited to a touch interface, such
as \cite{SM13, van2014finger}. Research by Sae-Bae et. al  \cite{SMI12, SMI12b}
showed that users can be uniquely identified from their
multi-touch gestures on multi-touch devices with high-probability.
However, just like text passwords, graphical and gesture password
alternatives authenticate users only at the time of login and they
do not address unauthorized access by an attacker after the user
initially logged on into the device.

In this context, continuous authentication or active
authentication \cite{Sim07, Aksari09, Niinuma10} mechanisms have
emerged as a very promising approach to alleviate the security
problems that stem from poor authentication technology. Here,
instead of authenticating a user at the time of login, the system
continuously monitors aspects of the user behavior biometrics in
order to maintain authentication after login. Some earlier work on
behavior biometrics based continuous authentication include
keystroke dynamics \cite{Clarke06, Serwadda13}, speaking pattern
\cite{Woo06} and device use patterns \cite{Furnell08, Li13}.

\begin{figure}
\centering
\includegraphics[height=2 cm,width=8 cm]{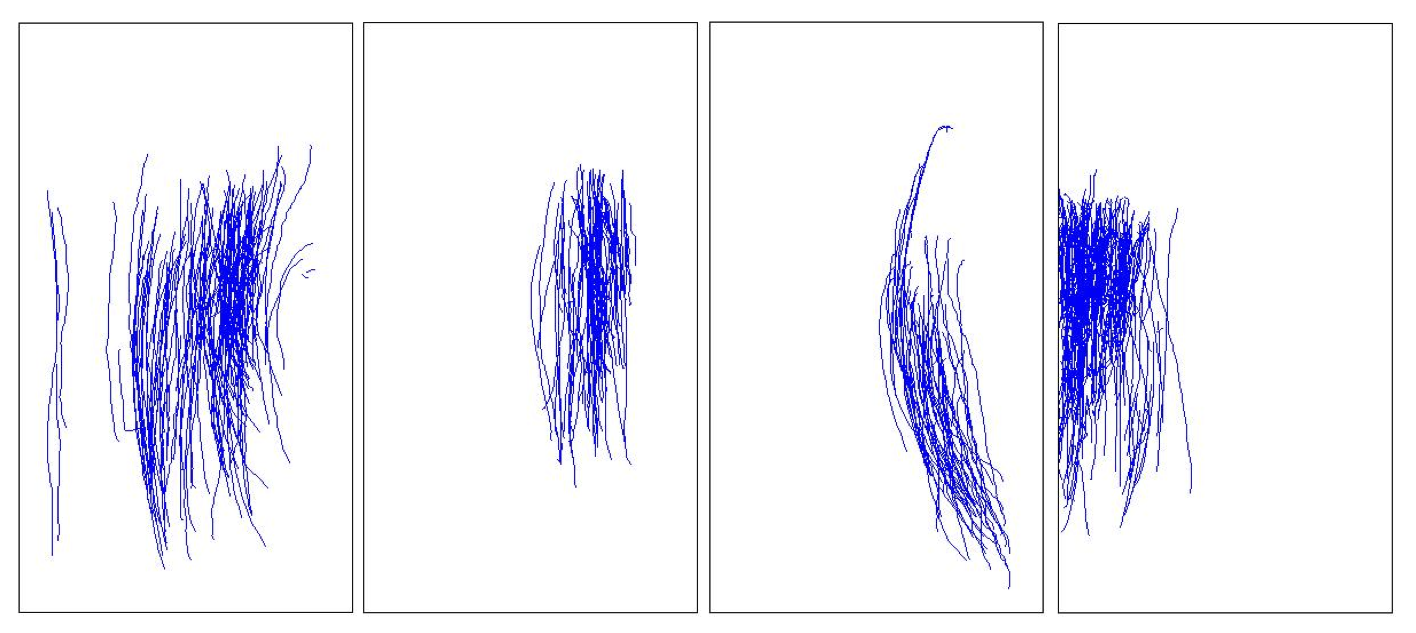}
\caption{Spatio-temporal touch patterns of four
users} \label{pic0}

\end{figure}

Based on touch behavior biometrics, a continuous authentication
method has been developed by Frank et. al. \cite{Frank13}
(Touchanalytics). They observed that during a stroke on the touch
screen of the mobile device, the spatio-temporal pattern (as shown
in Figure \ref{pic0}) of fingers along with the area of touch and
pressure is quite distinctive for every person. They reported high
performance when using multiple movements to authenticate the
users. Based on this observation various systems have been
developed, like, SenGuard \cite{Senguard11}, FAST \cite{FAST12},
and SilentSense \cite{SilentSense13}. Most of them used other
modalities along with touch behavior biometrics, like, motion,
voice, location history, walking pattern, to increase accuracy.

However, the classifiers (k-Nearest Neighbor, Support Vector
Machine) employed by the above mentioned approaches including
\cite{Frank13} require training data from both the owner as well
as other users for training. Since obtaining training data from
other users is not feasible, an authentication method that does
not need data from other users during training is desirable.

This paper presents an HMM based algorithm for continuous
authentication on touch devices which serves this purpose. This
research is built on the premise that to implement continuous
authentication in a feasible way, a method which offers the
possibility of being trained with only users data and can be
updated with new data over a period of time is needed. HMM can be
trained and updated with time, and is also relatively simple and
feasible to use, which makes it a good choice for continuous
authentication.

The key contributions of this paper are twofold. First, an HMM
based behavior model from the owner's touch information is
developed. Second, an in-depth analysis of the proposed method is
carried out in different usage scenarios.

The proposed HMM system is tested and compared to the performance
of the Touchanalytics system \cite{Frank13}. The rest of the paper
is organized as follows. Section 2 presents the proposed approach
in detail. In Section 3 the extensive experimental results are
discussed and finally Section 4 concludes the paper. 

\section{Proposed Approach}

The proposed framework works in two steps: training and
authentication. During training, a behavior model is created based
on the horizontal or vertical touch behavior (pressure, area,
duration and position) of a user using HMM. At the time of
authentication, the test observations are compared with the stored
behavior model to establish the identity of the user.

As mentioned in Section 1, HMM is considered for modeling the
stroke patterns of a subject since it is able to capture the local
dynamic characteristics of a stroke as well as its shape and
length. The touch pattern of a subject is modeled by a double
stochastic process, characterized by a given number of states each
of which is modeled by a mixture of Gaussians. The left-right
topology is chosen with no state skip allowed since it can
efficiently describe continuous processes. HMM allows modeling of
temporal variations, where the duration of the state
is variable.
The states then capture the transitive properties of the
consecutive coordinates of the stroke. Thus, the state transition
matrix represents the dynamic properties of the strokes. The state
sequence that maximizes the probability of observing the training
strokes becomes the corresponding model of a subject.

\subsection{Training HMM}
After normalizing the data set, it is used for training the HMMs
and finding the optimum parameters. As a first step, the state
transition matrix is initialized and the prior probability matrix
by random variables without making any assumptions on the touch
patterns. Then, training of HMM is
done from the initial set of strokes of a subject. The optimum number of states and mixtures of an HMM depend on the
complexity and average length of strokes in the training sequences
and their inter-variations. To provide sufficient evidence to
every Gaussian of every state in the training stage, the number of
mixtures times the number of states should be much smaller than
the length of the strokes. The Baum-Welch algorithm~\cite{Rabiner93} has been employed for estimating the HMM parameters for each subject.
Five-fold cross validation principle is used to estimate the
optimal number of states and the associated HMM parameters. Since
the parameters yielding the highest likelihood on the validation
set has been chosen, the model conveniently characterizes the
distinct stroke patterns for each subject while avoids
over-fitting.

\subsection{Authentication using HMM}

Once the behavioral models for all subject classes have been
learned through HMMs, authentication of the subjects can be
performed by computing the log-likelihood of the input strokes
using the Viterbi algorithm \cite{Rabiner93}. Since the length of
the stroke influences the log-likelihood (the log-likelihood
decreases exponentially with the increase of the stroke length),
the latter is normalized by the stroke length.

However, since the normalized log-likelihood is length-invariant,
two strokes, one being a part of the other, may produce similar
normalized log-likelihood despite being of different lengths. So,
an additional measure named as \emph{stroke
kinematics} is introduced. It represents the percentage of time spent in each
state. Since states represent segments of atomic motions between
points of change in motion pattern, the stroke kinematics captures
the detail dynamic properties of the strokes. The same
Viterbi algorithm \cite{Rabiner93} is used to compute the most likely
path. Then, if there are N states in the claimed identity's HMM,
stroke kinematics is computed as an N-component vector where the
$i^{th}$ component represents the fraction of time spent in the
$i^{th}$ state. Next, the similarity scores derived
from the normalized log-likelihood value and the stroke kinematics
for authentication are described.

\subsubsection{Similarity Score Computation}

\textbf{Likelihood Score:} The likelihood distance $D_{l}$ between
the normalized log-likelihood of the test stroke $L_t$ and the
average log-likelihood $L_a$ of the training database is
calculated as: $D_l = L_a - L_t$. Then the Likelihood score
$S_{l}$ is computed as follows: $S_{l} = exp\frac{-D_{l}}{P}$,
where $P$ is the number of touch features.

\textbf{Kinematic Score:} The stroke kinematics $SK_i$ for each of
the training strokes $i$ are computed beforehand. Then, for a test
stroke, the Euclidean distance $D_{e}^{i}$ between its
stroke kinematics ($SK_t$) and all the stroke kinematics of the
training database ($ D_{e}^{i} = \| SK_i - SK_t \| \forall i$) is calculated.
Next, the average of

these
distances ($\mathbb{D}$) is computed as: $\mathbb{D} = \frac{1}{M}
\sum_{i = 1}^{M} D_{e}^{i}$, where $M$ is the size of the training
data set. This average distance $\mathbb{D}$ is then used to
compute the Kinematic Score $S_{k}$ by an exponential function:
$S_{k} = exp\frac{-\mathbb{D}}{Q*N} $. The normalization factor
$Q$ in the denominator corresponds to the number of Gaussian
mixtures and $N$ is the number of components of the stroke
kinematics.

After getting the two similarity measures ($S_l$ and $S_k$) of a
test stroke, these two scores are combined by taking simple
arithmetic mean. The combined similarity score ($S_c$) is used for
final authentication.

\subsubsection{Multiple Strokes Fusion}

Since authentication using single stroke is highly volatile, to
increase the robustness of the authentication method, multiple
consecutive strokes are used for the final decision. The average
of all the combined similarity scores $S_c$, obtained from a
sequence of strokes, is employed for this purpose.

\section{Experimental Results and Discussion}

The authentication system is evaluated through calculation of
\emph{False Acceptance Rate (FAR)} and \emph{False Rejection Rate
(FRR)}. Since these two error rates are inversely related (lower
FAR increases the system security while lower FRR increases its
usability) \emph{Equal Error Rate (EER)} is also measured, where
FAR is equal to the FRR value.

\subsection{Data Set Description}

In absence of any other public touch databases, the data set of
Frank et al. \cite{Frank13} was chosen for its varied test
scenarios and realistic nature. Scrolling and horizontal stroke
data sets were collected from 41 subjects using four Android
phones with similar specification. More details about the data set
can be found in \cite{Frank13}. Based on this data, experiments
were designed to analyze three different application scenarios
with increasing problem difficulty, namely, short-term,
inter-session and long-term authentication. The same experimental
setup was followed to compare this approach with \cite{Frank13}.
The experimental results of the proposed approach in each of the
three situations are described in the following subsections.

\subsection{Short-term or Intra-session Authentication}

Short-term authentication is carried out to check whether the
authorized user is actually using the phone after successful
login. Therefore, authentication is done during the same session
of interaction. The training data set was created by randomly
drawing data from all available sessions of the two days and the
remaining data were used for testing.

\textbf{EER Performance:} Since training and testing is done in
the same session, authentication in this case is less challenging.
For single stroke, the median EER is found to be 5.35\% for
horizontal stroke HMM and 5.63\% for scrolling stroke HMM.

When computing the performance for 11 strokes (similar to \cite
{Frank13}), the EER of the proposed method decreases to 0.43\% and
0.31\% for horizontal and scrolling HMMs

respectively (see Table \ref{tab1}).

\begin{table}
\caption{Median EER rates - the newly proposed HMM-based approach
reaches low EER levels for all scenarios} \label{tab1} \centering
\small
\begin{tabular}{|c| p{.65 cm}| p{.8 cm}| p{.65 cm}| p{.65 cm}| p{.8 cm}| p{.65 cm}|}
\hline
\cline{2-7} \multicolumn{1}{|c||}{No.} & \multicolumn{6}{c|}{Equal Error Rate EER (\%)} \\
\cline{2-7} \multicolumn{1}{|c||}{of} & \multicolumn{3}{c|}{Horizontal HMM} & \multicolumn{3}{c|}{Scrolling HMM} \\
\cline{2-7}\multicolumn{1}{|c||}{Strokes} & {Short-term} & {Inter-session} & {Long-term} & {Short-term} & {Inter-session} & {Long-term}\\
\cline{1-7} \multicolumn{1}{|c||}{1} & {5.35} & {7.42} & {9.91} & {5.63} & {8.17} & {8.51}\\
\cline{1-7} \multicolumn{1}{|c||}{11} & {0.43} & {1.88} & {1.75} & {0.31} & {1.53} & {2.80}\\
\hline
\end{tabular}

\end{table}

\textbf{FAR and FRR Performance:} Application where security is
not so much of importance (like

games),
low FRR is desired.

When
the HMM system FRR is zero,
the median FAR is 6.78\% using one stroke for scrolling HMM and
7.13\% for horizontal HMM.
After observing 11 strokes, the FAR is reduced to 0.17\% and
0.54\% for scrolling and horizontal HMMs, respectively.

This shows that the
proposed method is highly secure even when FRR is zero, i.e. most
usable.

For application where high security is required (like banking),
FRR performance of the HMM algorithm is evaluated keeping zero
FAR. For one stroke, the median FRR was found to be 19.19\% for
horizontal HMM and 18.28\% for scrolling HMM. The FRR becomes
0.97\% after observing 11 strokes for horizontal HMM and 1.65\%
for scrolling HMM. The results indicate that in highest security
situation (FAR = 0), the usability of the proposed method is quite
high ($<$2\% FRR).

Since training and testing is done in the same session, the short
term test case is more of a `proof of concept' and less
challenging. If the attacker gets hold of the device just after
successful login, the short-term behavior model will be built from
the data of the attacker (since there are no data of the user in
the current session and the short-term model does not consider
owner's data from previous sessions). Thus, the attacker will be
recognized as the legal one for all further interactions in that
session. Therefore, the short-term classifier does not depict
realistic situation. The inter-session and long-term or inter-week
sessions represent more feasible scenarios since the owner's data
from previous sessions are considered during authentication.

\subsection{Inter-session Authentication}

In inter-session authentication, the user is authenticated across
multiple sessions with a brief time gap. Continuous authentication
in such scenario would enable the user to use the phone seamlessly
without unlocking each time after short burst of activity.

In this case, there was a time gap (of 10-12 minutes) between the
initial training session and the following two testing sessions.

\begin{figure}
\centering
\includegraphics[height=4 cm,width=8 cm]{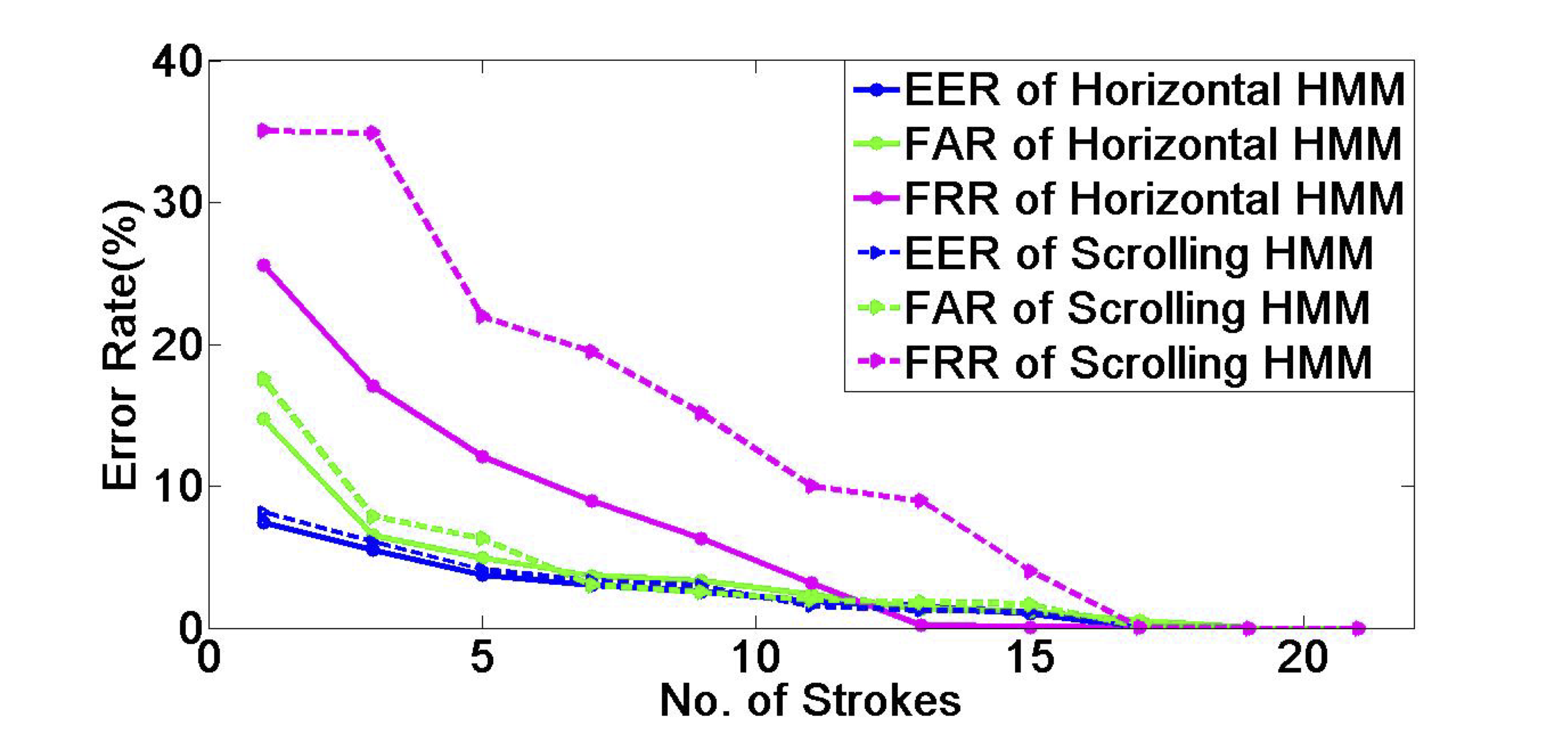}
\caption{Error rate variation as a function of
the number of strokes during inter-session authentication}
\label{pic2} 
\end{figure}

\textbf{EER Performance:} The EER performance variation with the
number of strokes is shown in blue lines in Figure \ref{pic2}.
Using single stroke, the median EER is 8.17\% for scrolling HMMs.

(see Table
\ref{tab1}).

The median EER is 1.88\% for horizontal and
1.53\% for scrolling HMM. The EER becomes zero after 17 horizontal
or scrolling strokes.

\textbf{FAR and FRR Performance:} The FAR performance of the
proposed approach is shown in green lines Figure \ref{pic2} while
FRR is zero. For 11 strokes, FAR is 1.83\% for scrolling HMM and
2.33\% for horizontal HMM. Similarly, keeping FAR value zero, the
variation of FRR with respect to stroke number is plotted in
magenta lines in Figure \ref{pic2}. After 11 strokes, FRR is found
to be 10.00\% for scrolling HMM and 3.20\% for horizontal HMM.

\subsection{Long-term Authentication}

Here the training set comprises of the data collected during
multiple sessions of the first day. Then, testing is done using
the data captured a week later. Thus, long-term authentication
tries to evaluate the classifier when the time gap between the
training and testing is quite high. Due to this time gap,
authentication in this case is the most challenging one.

\begin{figure}
\centering
\includegraphics[height=4 cm,width=8 cm]{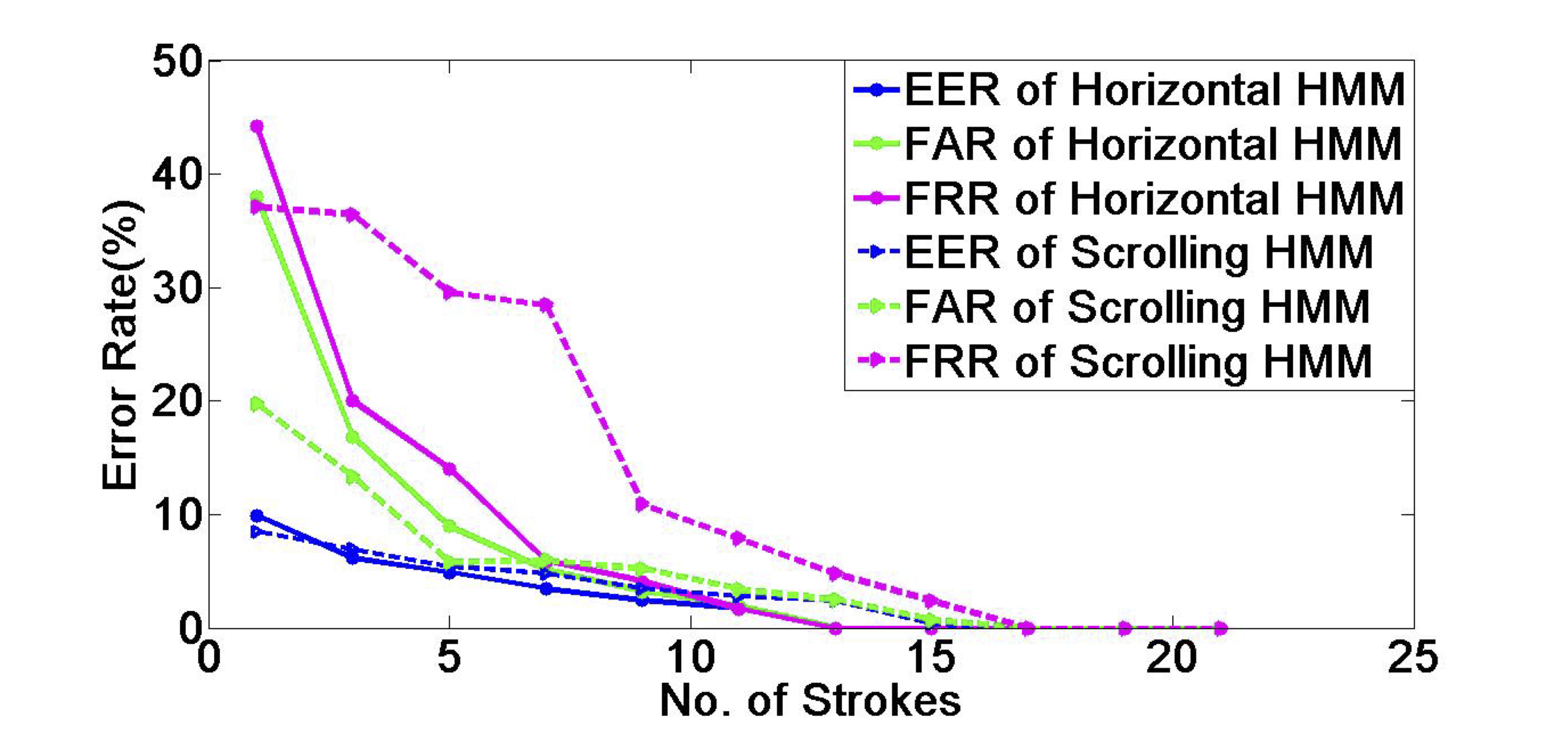}
\caption{Error rate variation as a function of
the number of strokes during long-term authentication}
\label{pic3} 
\end{figure}

\textbf{EER Performance:} The EER performance variation with the
number of strokes is shown in blue lines in Figure \ref{pic3}. For
single stroke, the median EER is found to be 9.91\% for horizontal
HMM and 8.51\% for scrolling HMM. Using 11 strokes, EER decreases
to 1.75\% for horizontal HMM and 2.8\% for the scrolling HMM.

\textbf{FAR and FRR Performance:} The green lines of Figure
\ref{pic3} plot the FAR performance of the proposed approach,
while FRR is zero. For 11 strokes, FAR is found to be 3.39\% for
scrolling HMM and 1.96\% for horizontal HMM. The FRR variation
with stroke number is plotted in magenta lines in Figure
\ref{pic3}, while FAR is zero. For 11 strokes, FRR is 7.9\% for
scrolling HMM and 1.69\% for horizontal HMM.

\begin{table}
\caption{Comparative results of the proposed approach against the
Touchanalytics approach using 11 strokes} \label{tab2} \centering
\small
\begin{tabular}{|c| p{1.1 cm}| p{1.1 cm}| p{1.1 cm}| p{1.1 cm}|}
\cline{1-5} \multicolumn{1}{|c||}{} & \multicolumn{2}{c|}{Touchalytics} & \multicolumn{2}{c|}{HMM-based} \\
\multicolumn{1}{|c||}{} & \multicolumn{2}{c|}{ \cite{Frank13}} & \multicolumn{2}{c|}{Approach} \\
\cline{2-5} \multicolumn{1}{|c||}{} & {Worst EER(\%)} & {Median EER(\%)} & {Worst EER(\%)} & {Median EER(\%)} \\
\cline{1-5} \multicolumn{1}{|c||}{Short-term} & {12} & {$\sim$0\%} & {5} & {$\sim$0\%}\\
\cline{1-5} \multicolumn{1}{|c||} {Inter-session} & {28} & {2-3\%} & {12} & {\textless 2\%} \\
\cline{1-5} \multicolumn{1}{|c||}{Long-term} & {38} & {\textless 4\%} & {28} & {\textless 3\%}\\
\hline
\end{tabular}

\end{table}

\subsection{Performance Comparison}

The results of the HMM algorithm are compared to the
Touchanalytics algorithm in all three scenarios, i.e., short-term,
inter-session and long-term, in Table \ref{tab2}. The HMM
algorithm performs better in all the test scenarios than
\cite{Frank13}. A special case is the long-term situation, which
is the most challenging one due to one week time gap between
training and testing. Since there are only 14 users' data for this
study, the HMM algorithm used all of them without classifying any
as outliers. Due to the small data set size, using all 14 users'
data is expected to provide more accurate results and therefore
these results are more indicative of the expected performance.
This is in contrast to the Touchanalytics, which classified the
worst 3-5 results as outliers and did not use them for computing
the overall median EER. Therefore, the results cannot be compared
directly (the Touchanalytics authentication achieves less than 4\%
for the long-term using only the best 9-11 users).

In addition, since the proposed HMM based approach trains the
model based on only the owner's data, it basically acts as one
class classifier to detect whether the current user is legitimate
or not. On the contrary, all the state-of-the-art approaches
including \cite{Frank13} employ binary classifiers that are
expected to give good results due to use of training data from the
owner as well as other users \cite{SilentSense13}.

Better performance of the proposed approach indicates inherent
strength of the HMM based behavior model.

\section{Conclusions}
This work introduces a new touch behavior modeling approach using HMM. Since HMM allows
automatic continuous training and data updating, it offers
significant advantage for continuous authentication. This work is
the first one that uses HMM for
continuous authentication based on mobile-phone user input.

The authentication method is based only on the stroke patterns
recorded from the owner's touch interactions on his mobile device.
Extensive evaluation of the proposed approach on the

Touchanalytics
database has been carried out.

The results of the HMM algorithm were compared to the
Touchanalytics algorithm and found to be superior, without using
any training data from other users. The benefits of using only the
device owner's data are twofold. First, in case of personal
devices, data from other users may not be available. Thus,
training the classifier with other users' data is not possible.
Second, authentication results with only owner's data reflect the
real-life situation in a better way.

This work also looks at the security and the usability of the
proposed approach (i.e., in cases where security is critical and
FAR$=$0, or when high usability is needed and FRR$=$0). The
results show that the approach has the potential to be used for
user authentication in continuous and implicit manner. Future work
involves more extensive evaluation of the approach with a newly
generated data set featuring other types of touch patterns and
sensory information.

\bibliographystyle{abbrv}

\begin{thebibliography}{10}


\bibitem{AD04}
S. Akula and V. Devisetty, ``Image based registration and
authentication system," in \emph{Midwest Instruction and Computing
Symposium}, vol. 4, 2004.
%

\bibitem{Aksari09}
Y. Aksari and H Artuner, ``Active authentication by mouse
movements,"\emph{ Computer and Information Sciences (ISCIS), 24th
International Symposium on,} IEEE, 2009.
%

\bibitem{Birget06}
J.C. Birget, D. Hong, and N. Memon, ``Graphical passwords based on
robust discretization," \emph{IEEE Transactions on Information
Forensics and Security,} vol. 1, no. 3, pp. 395--399, 2006.

%


\bibitem{SilentSense13}
C. Bo, L. Zhang, and X.-Y. Li,``SilentSense: Silent user
identification via dynamics of touch and movement behavioral
biometrics," arXiv preprint arXiv:1309.0073, 2013.

\bibitem{Clarke06}
N.L. Clarke and S.M. Furnell, ``Authenticating mobile phone users
using keystroke analysis," \emph{International Journal of
Information Security}, vol. 6, no. 1, pp. 1--14, 2006.
%

\bibitem{data12}
DataGenetics. PIN Analysis. Online, http://www.datagenetics.
com/blog/september32012/; October 2013.


\bibitem{DMR04}
D. Davis, F. Monrose, and M. Reiter, ``On user choice in graphical
password schemes", in \emph{13th Usenix Security Symposium,} 2004,
vol. 13, pp. 1--14.


\bibitem{DP00}
R. Dhamija and A. Perrig, ``Dej�a Vu: User study using images for
authentication," \emph{9th Usenix Security Symposium}, 2000, pp.
14--17.


\bibitem{DMB07}
A.E. Dirik, N. Memon, and J.C. Birget, ``Modeling user choice in
the PassPoints graphical password scheme," in \emph{Proc. of the
3rd symposium on Usable privacy and security (SOUPS)}, ACM, 2007,
pp. 20--28.
%

\bibitem{FAST12}
T. Feng, Z. Liu, K.-A. Kwon, W. Shi, B. Carbunar, Y. Jiang, and N.
Nguyen, ``Continuous mobile authentication using touchscreen
gestures," in \emph{Homeland Security (HST), 2012 IEEE Conference
on Technologies for}, IEEE, 2012, pp. 451--456.
%
%
\bibitem{Frank13}
M. Frank, R. Biedert, E. Ma, I. Martinovic, and D. Song,
``Touchalytics: On the applicability of touchscreen input as a
behavioral biometric for continuous authentication," \emph{IEEE
Trans. on Information Forensics and Security}, vol. 8, no. 1, pp.
136--148, 2013.
%

\bibitem{Furnell08}
S. Furnell, N. Clarke, and S. Karatzouni, ``Beyond the pin:
Enhancing user authentication for mobile devices," \emph{Computer
Fraud and Security}, vol. 2008, no. 8, pp. 12--17, 2008.
%

\bibitem{Jo12}
H.H. Jo, M. Karsai, J. Kerte`sz, and K. Kaski, ``Circadian
patterns and Burstiness in mobile phone Communication," \emph{New
Journal of Physics}, vol. 14, 2012.

\bibitem{Li13}
F. Li, N. Clarke, M. Papadaki, and P. Dowland, ``Active
authentication for mobile devices utilising behaviour profiling,"
\emph{International Journal of Information Security}, pp. 1--16,
2013.
%

\bibitem{Niinuma10}
K. Niinuma, U. Park, and A.K. Jain, ``Soft biometric traits for
continuous user authentication," \emph{IEEE Transactions on
Information Forensics and Security,} vol. 5, no. 4, pp. 771--780,
2010.


%
\bibitem{Rabiner93}
L. Rabiner and B.H. Juang, ``Fundamentals of speech recognition,"
\emph{Prentice Hall Signal Processing Series}, 1993.


\bibitem{SMI12}
N. Sae-Bae, N. Memon, and K. Isbister,``Investigating multi-touch
gestures as a novel biometric modality," in \emph{Biometrics:
Theory, Applications and Systems (BTAS), 2012 IEEE Fifth
International Conference on}, IEEE, 2012, pp. 156--161.
%


\bibitem{SM13}
N. Sae-Bae and N. Memon,``A simple and effective method for online
signature verification," in \emph{Biometrics Special Interest
Group (BIOSIG), 2013 International Conference of the}, IEEE, 2013,
pp. 1--12.
%


\bibitem{SMI12b}
N. Sae-Bae, N. Memon, and K.Isbister,``Biometric-rich gestures: A
novel approach to authentication on multi-touch devices," in
\emph{Proc. of the 2012 ACM annual conference on human factors in
computing systems}, ACM, 2012, pp. 977--986.


\bibitem{Serwadda13}
A. Serwadda, Z. Wang, P. Koch, S. Govindarajan, R. Pokala, A.
Goodkind, D.-G. Brizan, A. Rosenberg, V.V. Phoha, K. Balagani,
``Scan-based evaluation of continuous keystroke authentication
systems," \emph{IT Professional}, vol. 15, no. 4, pp. 20--23,
2013.


\bibitem{Senguard11}
W. Shi, J. Yang, Y. Jiang, F. Yang and Y. Xiong, ``Senguard:
Passive user identification on smartphones using multiple
sensors," in \emph{Wireless and Mobile Computing, Networking and
Communications(WiMob), 2011 IEEE 7th International Conference on},
IEEE, 2011, pp. 141--148.


\bibitem{Sim07}
T. Sim, S. Zhang, R. Janakiraman, and S. Kumar, ``Continuous
verification using multimodal biometrics," \emph{IEEE Trans.
Pattern Analysis and Machine Intelligence,} vol. 29, no. 4, pp.
687--700, 2007.

\bibitem{van2014finger}
T.~Van~Nguyen, N.~Sae-Bae, and N.~Memon.
\newblock Finger-drawn pin authentication on touch devices.
\newblock In {\em Image Processing (ICIP), 2014 IEEE International Conference
  on}, pages 5002--5006. IEEE, 2014.
  
\bibitem{roy2014hmm}
A.~Roy, T.~Halevi, and N.~Memon.
\newblock An hmm-based behavior modeling approach for continuous mobile
  authentication.
\newblock In {\em Proc. of IEEE Intl' Conf. on Acoustics, Speech and Signal
  Processing (ICASSP)}, pages 3789--3793. IEEE, 2014.

\bibitem{WK04}
D. Weinshall and S. Kirkpatrick, ``Passwords you'll never forget,
but can't recall," in \emph{Conference on Human Factors in
Computing Systems (CHI)}, ACM, 2004, pp. 1399--1402.

\bibitem{Woo06}
R. Woo, A. Park and T. Hazen, ``The MIT mobile device speaker
verification corpus: data collection and preliminary experiments,"
in \emph{Speaker and Language Recognition Workshop, IEEE Odyssey
2006: The}, IEEE, 2006, pp. 1--6.

\end{thebibliography}

\end{document}